\documentclass{elsart5p}

\usepackage{amssymb}
\usepackage{graphicx}
\usepackage{dcolumn}
\usepackage{bm}
\usepackage{epsfig}

\begin{document}

\begin{frontmatter}



\title{Time-Evolution of a Fractal Distribution: Particle Concentrations in Free-Surface Turbulence}


\author{Jason Larkin}
\address{Department of Mechanical Engineering and Material Science, University of Pittsburgh, Pittsburgh, PA 15260, USA.}
\author{Walter Goldburg}
\address{Department of Physics \& Astronomy, University of Pittsburgh, Pittsburgh, PA 15260, USA.}
\author{M. M. Bandi}
\address{Center for Nonlinear Studies (T-CNLS) and Condensed Matter \& Thermal Physics Group (MPA-10), Los Alamos National Laboratory, Los Alamos, NM 87545, USA.}

\begin{abstract}
Steady-state turbulence is generated in a tank of water and the trajectories of particles forming a compressible system on the surface are tracked in time. The initial uniformly distributed floating particles coagulate and form a fractal distribution, a rare manifestation of a fractal object observable in real-space. The surface pattern reaches a steady state in approximately 1 s. Measurements are made of the fractal dimensions $D_q(t)$ ($q=1$ to $6$) of the floating particles starting with the uniform distribution $D_q(0)$ = 2 for Taylor Microscale Reynolds number $Re_{\lambda} \simeq 160$.   Focus is on the the time-evolution of the correlation dimension $D_2(t)$ as the steady state is approached. This steady state is reached in several large eddy turnover times and does so at an exponential rate.
\end{abstract}

\end{frontmatter}

\section{Introduction}
When studying a dynamical system in phase space, one of several scenarios may occur. The simplest cases involve either a fixed point or a limit cycle (periodic orbit). The most interesting scenario unfolds when studying dissipative systems undergoing chaotic evolution. When a system is driven out of thermal equilibrium, a phase point which previously would have visited all regions of phase space with almost equal probability, now spends most of its time in a limited region. There it develops a complex fractal topology with a non-trivial fractal dimension \cite{dorfmanbook}. Fractal dimensions have been measured extensively for simple maps (eg. the Henon map) \cite{ottbook, schusterbook} as well as for real world systems \cite{brandstater1983}. Typically, the fractal behavior of such systems is studied in an asymptotic limit which ignores the transient behavior \cite{ottbook}. In this paper, we show that measuring a fractal dimension in the transient state is an effective way to characterize the evolving topology of this particular system.

In this experiment we study the transient evolution of a fractal topology in a laboratory setting, namely, the clustering of floating particles at the surface of a turbulent fluid. If neutrally bouyant, non-inertial particles are introduced into a turbulent flow, they quickly distribute themselves throughout the volume of the fluid; further stirring leaves the particle density distribution uniform. The bouyant particles used in this experiment have a density much less than that of the fluid on which they move. Hence their behavior is entirely different.  Now, their motion is confined to the surface of the turbulent fluid.  If the spatial distribution is initially uniform, at say, $t$ = 0, continuous stirring of the underlying fluid evolves this spatial distribution into string-like structures.  Ultimately a steady state is reached, at which time the floaters occupy a fractal dimension much less than 2 \cite{Yu1990}.  This time evolution into a fractal is a generic effect. It occurs even if the interactions between the floaters is negligibly small.  A common manifestation of this phenomenon is the coagulation of scum on the surface of the sea, as is often seen in an ocean harbor.

The transient evolution of the floaters is studied by uniformly distributing the floaters at $t=0$ (as discussed in 2.). At times $t>0$ the particles are subject to the underlying turbulent flow until their spatial distribution reaches a steady state. To study the floater's fractal distribution, the spectrum of fractal dimensions $D_q(t)$ is measured at subsequent times $t$ during the transient evolution. One may argue that the generalized fractal dimensions $D_q$ of any system are meaningful only in the limit of the evolution time $t \rightarrow \infty$. However, there exist instances of fractal objects realizable in real or configuration space (such as shear flows \cite{Sreenivasan1989}, cement gels \cite{Pintar1988}, and \cite{Sommerer1996}) that are amenable to investigation of their transient behavior.

It is essential to realize that the floaters in this experment are passively advected by the underlying flow. They are small enough to follow the velocity field of the turbulent sea on which they move, in a plane that has coordinates $x$, $y$, $z=0$. There are, of course, waves on the surface, and they can drive the motion of the floaters. Separate studies have shown that the amplitude of the surface waves is small enough to have a negligible effect on the particle motion \cite{Cressman2004}.  Though the water molecules can have a downward velocity component at all values of the depth $z$, the floaters cannot follow them.  That is why they coagulate and disperse in the plane of their motion $x,y$ with $z=0$.  In this sense the floaters form a strongly compressible system. Using the following definition of the dimensionless compressibility,
 \begin{equation} \label{C}
 {\cal C}=\left\langle \left(\stackrel{\rightharpoonup}{\nabla }_{2} \cdot \stackrel{\rightharpoonup}{v}\right)^{2} \right\rangle / \left\langle \left(\stackrel{\rightharpoonup}{\nabla }_{2} \stackrel{\rightharpoonup}{v}\right)^{2} \right\rangle
 \end{equation}
With this definition, ${\cal C}$ must lie between zero (incompressible) and 1 (potential) for an isotropic flow field.  Experimentally, ${\cal C}$ is close to 0.5 \cite{Cressman2004}.

The coagulation phenomenon described above was demonstrated and analyzed by  J. Sommerer and E. Ott (S\&O) \cite{Ott1993}. The solution was rather gently stirred via a pulsing jet, its motion being slow enough that they could measure the steady-state fractal dimensions of the surface particles and the two Lyapunov exponents as well.  The latter parameters, $\lambda_1 > \lambda_2$ define the rate at which initially close particle pairs separate in time. Since the pattern becomes string-like in the steady state, the largest exponent $\lambda_1$ is positive, and the other one $\lambda_2$ is negative. Since the total area ultimately occupied by the floaters decreases, $\lambda_1 +\lambda_2 <0$. The information dimension ($D_1$) of the fractal pattern can be related to the dynamics of the system through the Kaplan-Yorke dimension $D_{KY}=1+\lambda_1/|\lambda_2|$ \cite{Sprott}. For a two dimensional system, $D_{KY}$=$D_1$. In a separate study \cite{Sommerer1996} (but with a similar experimental setup as in  \cite{Ott1993}), the correlation dimension $D_2$ of the floating particles was measured, both at an initial time $t \simeq  0$ when $D_2 \simeq 2$ and in the steady-state where $1 < D_2 < 2$. In that experiment, the correlation dimension could not be measured during the transient state because of poor scaling \cite{Sommerer1996}. In the present experiment, we observe a robust scaling of the correlation sum $C_2(r)$ during the transient evolution, see Figure \ref{D2_scaling}, from an initial state $D_2(0) \simeq 2$ to a steady-state value $D_2(\infty) \simeq 1.25$.

The present experiment differs from that of S\&O in that the stirred fluid, water, could be driven into a strongly  turbulent state. The Taylor microscale Reynolds number,  $Re_{\lambda}$ is approximately $160$ (see Table I). This large $Re_{\lambda}$ establishes a well defined inertial range of the flow (see \cite{Cressman2004}). For a more thorough exploration of this type of flow, see \cite{Cressman2004,Goldburg2001}. Since the experiment of S\&O, there have been several theoretical advances pertaining to clustering phenomena in turbulent flows. These theories utilize the statistics of stretching rates below the dissipative scale $\eta$ of turbulence (Table I) to predict a multi-fractal particle distribution for compressible flows \cite{Bec2004}. The measurements made in this experiment are strictly for scales greater than the dissipative scale $\eta$ (see section 2), where there is currently no theoretical guidance \cite{Bec2004,Balkovsky2001}.

\begin{figure}[h]
\begin{center}$
\begin{array}{cc}
\includegraphics[width=3.25 in]{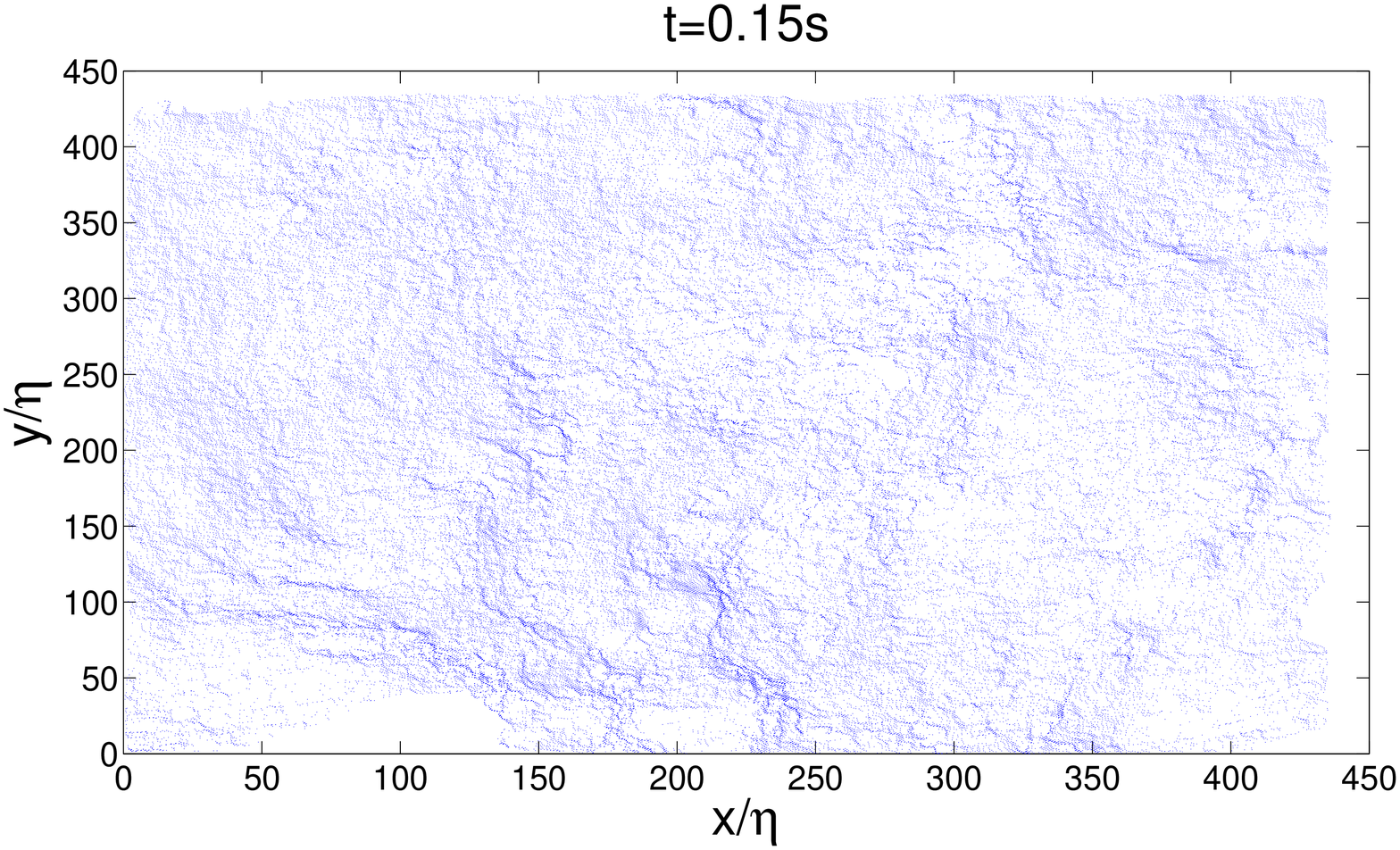} &\\
\includegraphics[width=3.25 in]{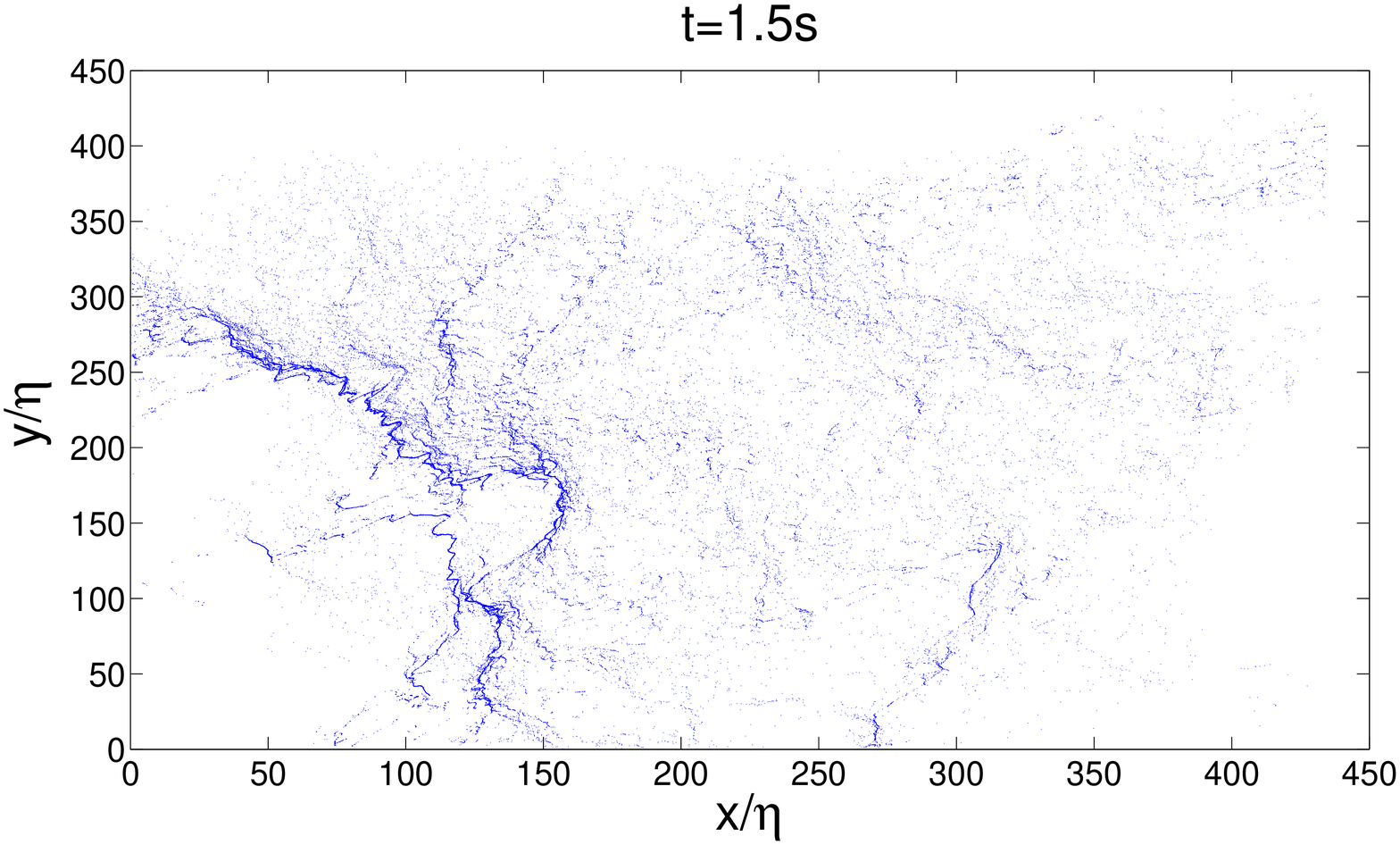}
\end{array}$
\end{center}
\caption{Visualization of particles clustering
from a uniform initial distribution of $10^5$ tracer particles for Re$_\lambda$=169 at t=0.15s for (top) and t=1.5s (bottom). By t=1.5s the particle distribution has reached a steady-state,  which occurs within several Large Eddy Turnover Times $\tau _{0}$ (Table \ref{table}.)}
\label{particles}
\end{figure}

Before describing the experiment in detail, it is helpful to observe the pattern of the floaters at a sufficiently long time so that the steady state has almost been achieved. Figure \ref{particles} shows the distribution of the Lagrangian particles (discussed below) at $t$ = 0.15 s and $t$ = 1.5 s, their spatial distribution being uniform at $t$ = 0. Were an image made at $t$ =0, the particle distribution (shown as dots) would be uniform. The blank white spaces are due to the finite initial seeding procedure and do not affect the results.

\section{Experiment}
\begin{figure}[h]
\begin{center}$
\begin{array}{cc}
\includegraphics[width=3.70 in]{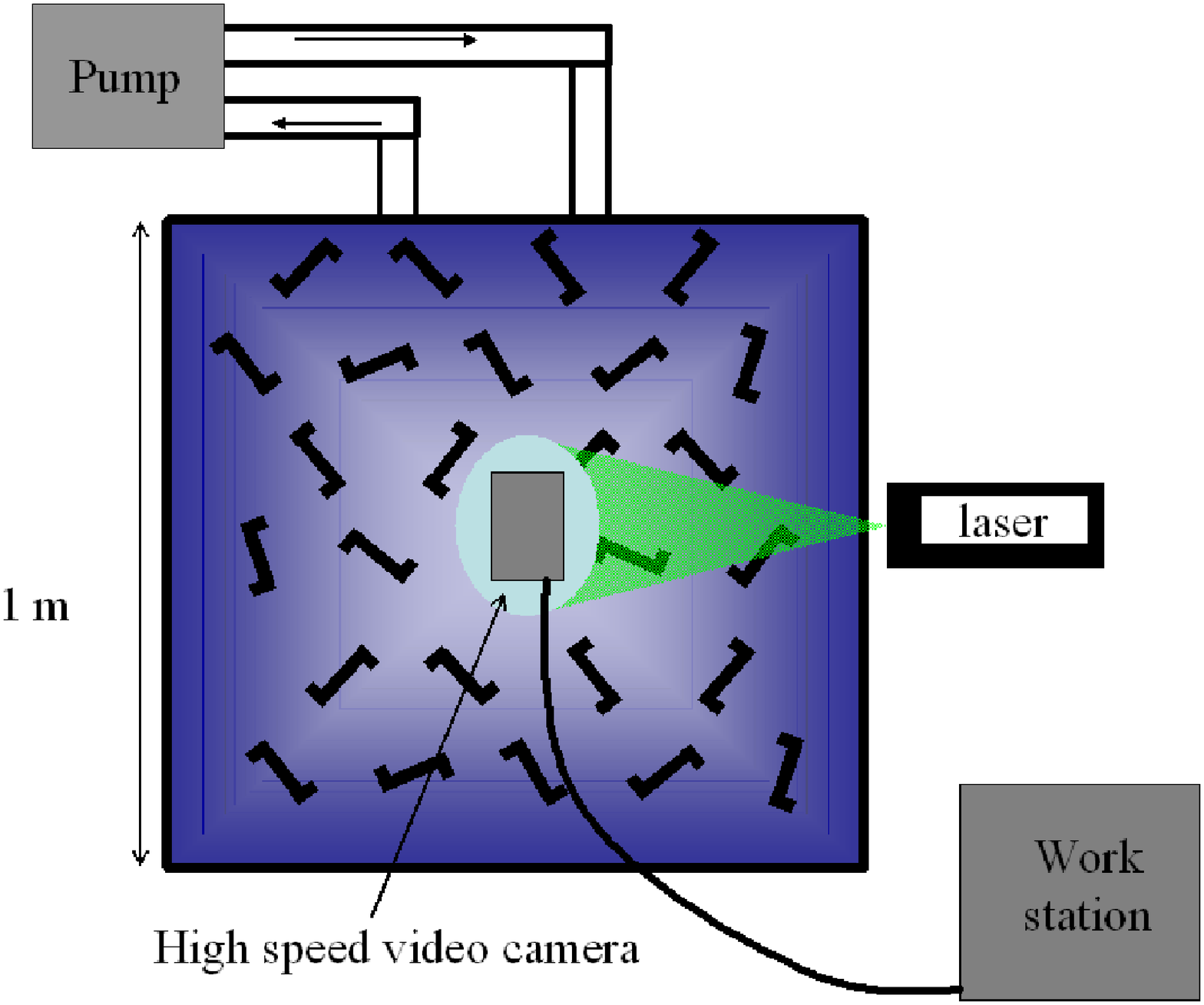} &\\
\includegraphics[width=3.70 in]{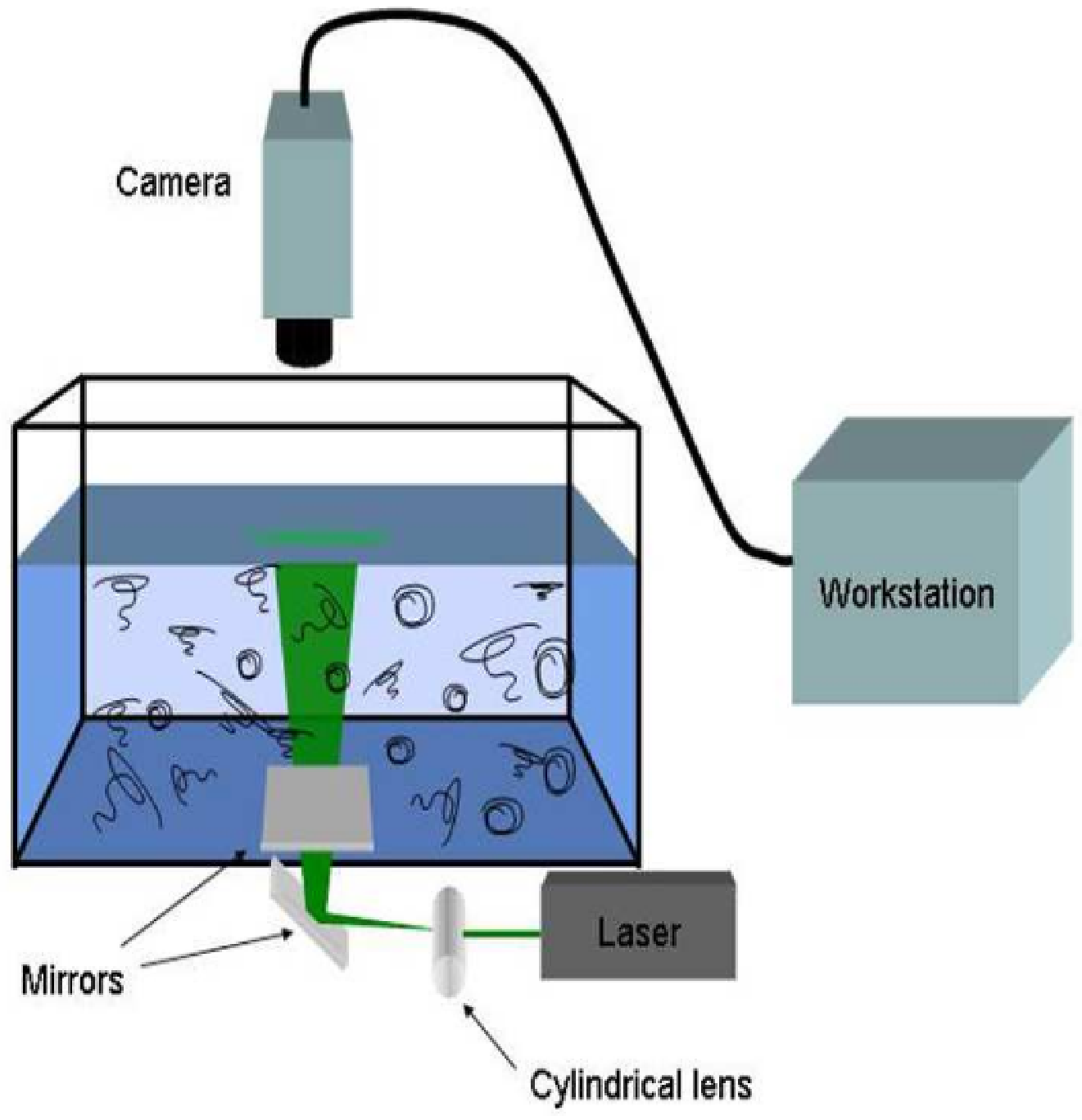}
\end{array}$
\end{center}
\caption{Schematic of the top-view (top panel) and side-view (bottom panel) of the experimental setup. 36 rotating capped jets are placed horizontally on the tank floor (shown as randomly oriented Z-shaped patterns) that pump water into the tank re-circulated by a 8hp pump. The central region of the water surface is illuminated by a laser-sheet. A high-speed digital camera suspended vertically above this central region captures images of the light scattered by buoyant particles (50 $\mu$m hollow-glass spheres of specific gravity 0.25).}
\label{setup}
\end{figure}

The 1 m $\times$ 1 m tank is filled with water to a height of $30 cm$. The tank is large compared to the camera's field of view.  The turbulence is generated by a large pump connected to a network of rotating jets in a plane 10 cm above the tank floor. See Figure \ref{setup} for a schematic of the experimental setup. The arrangement creates uniform turbulence in the center of the tank, and also moves the source of turbulent injection far from the fluid surface where the measurements are made \cite{Cressman2004}.  With this scheme, surface waves, which cannot be avoided,  do not exceed an amplitude of $\sim$1 mm \cite{Cressman2004}.  It is necessary that the surface of the tank be freshly cleaned before each set of measurements. Otherwise, amphiphiles form a continuous layer on the surface and prevents the floaters from moving freely under the action of the turbulence \cite{Cressman2004}.

The hydrophillic particles chosen here are subject to capillary forces which
are very small compared to forces coming from the turbulence, and do not affect the results as they do in \cite{PFL2006, FWDL2005}. The non-inertial character of the particles is minimal because the Stokes number $St$
is small: $St=\tau_s v_{rms}/a \simeq 0.01$, where $a$ is the particle radius,
$v_{rms}$ is the RMS velocity of the turbulent fluid at the free-surface, and
$\tau_s$ is the stopping time of the particle \cite{BCG2008}.

During an experimental run, the floating particles ($ 50 \mu m$ diameter and specific gravity of 0.25) are constantly seeded into the fluid from the tank floor, where they undergo turbulent mixing as they rise due to buoyancy and are uniformly
dispersed by the time they rise to the surface. Once at the free-surface,
their motion is constrained to the two-dimensional surface plane. Their motion is tracked via a high-speed camera (Phantom v.5) situated above the tank. The camera field-of-view is a square area of side length $L = 9 cm$. The constant particle injection is necessary to replace particles at the surface during the
experiment. The source and sink structure at the surface fluctuates in both time and space, which can cause particles to leave the camera's field of view.

Instantaneous velocity fields are measured using an in-house developed particle imaging velocimetry (PIV) program which processes the recorded images of the floaters. The constant injection of particles ensures that surface sources and sinks receive an adequate
coverage of particles on the surface. The local particle density at the surface determines the average spacing of the velocity vector fields produced by the PIV program. The resulting velocity vectors are spaced (on average) by $\delta x =$ 2.5 $\eta$ over both sources and sinks. This vector grid spacing is important for the Lagrangian particle evolution scheme, which is discussed below. The camera's height above the water surface was chosen so that a pixel size is
roughly $0.1 mm$, comparable to the dissipative scale
of the turbulence.

The measured velocity field was then used to
solve the equation of motion for Lagrangian particles : \begin{equation}
\label{advection}
\frac{d{\bf x}_i}{dt} = {\bf v}({\bf x}_i(t),t),
\end{equation}
where ${\bf v}({\bf x}_i,t)$ is the velocity field and ${\bf x_i}=(x_i,y_i)$ are the individual particle positions. To achieve accurate results for the Lagrangian particle evolution, the vector fields used in Eq. \ref{advection} were interpolated from the experimentally determined velocity vectors via a
bi-cubic interpolation scheme developed for numerical simulations, as discussed in \cite{Pope1988} and implemented in \cite{Cressman2004}. This scheme uses the smooth flow between grid points separated by length scales comparable to $\eta$ to interpolate the velocity field between measured velocity grid points.  To use this scheme it is necessary for the measured velocity grid spacing to satisfy the criterion $\delta x < \pi \eta$, where $\delta x$ is the above mentioned average measured velocity grid spacing. We have tested to ensure that the results do not depend on the velocity grid spacing by varying the spacing from $\delta x$ = 2.5 $\eta$ to $\delta x$ = 4 $\eta$.

The Lagrangian particle tracks evolved by Eq. (\ref{advection}) are then used for the measurements presented in this work. This is the method used to achieve a uniform distribution of floaters at $t=0$. Visualization of these Lagrangian tracers can be seen in Figure \ref{particles}. The experimental setup is discussed in more detail in \cite{Cressman2004}.  Data were taken for several values of $Re_\lambda \simeq 150-170$ with an average $Re_\lambda \simeq 160$. Since the measurements show no systematic variation with the $Re_\lambda$ over this range, each experimental run was averaged to decrease measurement errors. Turbulent parameters measured at the surface are listed in Table \ref{table}.
All of the statistics presented below were obtained by evolving
$\sim 10^5$ Lagrangian particles per frame. Tests were performed to ensure that the number of tracers provided adequate statistics to calculate $D_q(t)$. The initial homogeneous seeding of particles was varied from $10^5$ to $4 \times 10^5$ and the results were insensitive to this variation.  Since computation time goes roughly as $N^2$ because of Eq. (\ref{EQ12}), no more  than $10^5$ tracers were used.

\begin{table}
\caption{\label{table}Turbulent parameters measured at the surface.  Measurements are made at several values of the $Re_\lambda$ with an average $Re_\lambda \simeq 160$.  The parameters listed are averages, with deviations less than $10\%$.}
\begin{tabular}{|p{1.5in}|p{1.2in}|p{0.7in}|} \hline
Parameter & Symbol used in text  & Measured Value \\ \hline
\small {Taylor microscale $\lambda $ (cm)}
\newline \newline \newline Taylor Re$_\lambda$
\newline \newline Integral Scale  $l_0$ (cm)
\newline \newline \newline Large Eddy Turnover time (LETT) $\tau _{0}$ (s)
\newline \newline \newline  Dissipation Rate $\varepsilon_{diss}$ (cm2/s3)
\newline \newline Kolmogorov scale $\eta $ (cm)\newline \newline RMS
Velocity vrms (cm/s)
\newline \newline \newline \newline Compressibility $C$

& $\lambda $=$\sqrt{\frac{v^{2} _{rms} }{\left\langle \left({\partial v_{x}
\mathord{\left/{\vphantom{\partial v_{x}  \partial
x}}\right.\kern-\nulldelimiterspace} \partial x} \right)^{2}
\right\rangle } }$
\newline \newline \newline Re$_\lambda$=$\frac{v_{rms} \lambda }{\nu } $
\newline \newline $l_0$=$\int dr\frac{\left\langle v_{\left\| \right. }
(x+r)v_{\left\| \right. } (x)\right\rangle }{\left\langle
\left(v_{\left\| \right. } (x)\right)^{2} \right\rangle }$
\newline \newline \newline $\tau _{0} =\frac{l_{0} }{v_{rms} } $
\newline \newline \newline $\varepsilon_{diss}$=$10\nu \left\langle \left(\frac{\partial v_{x} }{\partial x} \right)^{2} \right\rangle $
\newline \newline \newline $\eta =\left(\frac{\nu ^{3} }{\varepsilon }
\right)^{1/4} $
\newline\newline $v_{rms} =\sqrt{\left\langle v^{2} \right\rangle-\left\langle v\right\rangle ^{2} }$
\newline \newline \newline ${\cal C}=\frac{\left\langle \left(\stackrel{\rightharpoonup}{\nabla }_{2} \cdot \stackrel{\rightharpoonup}{v}\right)^{2} \right\rangle }{\left\langle \left(\stackrel{\rightharpoonup}{\nabla }_{2} \stackrel{\rightharpoonup}{v}\right)^{2} \right\rangle }$.

& \small{0.47
\newline \newline \newline 160
\newline \newline 1.42
\newline \newline \newline 0.43
\newline \newline \newline 6.05
\newline \newline \newline 0.02
\newline \newline \newline 3.3
\newline \newline \newline 0.49 $\pm$ 2\%}
\\ \hline
\end{tabular}
\end{table}

\section{Results and Analysis}
We investigate the inhomogeneous particle distributions by measuring their time-evolving fractal dimensions. Fractal dimensions are mathematical representations of complex patterns and provide
measures of spatial (or temporal) dependence at a variety of scales. For any $q$, the spectrum of fractal dimensions is \cite{Hilborn}:
\begin{equation} \label{EQ13}
D_{q} =\mathop{\lim }\limits_{r\to 0} \frac{1}{q-1} \frac{d\log (C_{q} (r))}{d\log r}
\end{equation}
where the correlation functions $C_q(r)$ are defined:
\begin{equation} \label{EQ12}
C_{q} (r)=\frac{1}{N} \sum _{i}^{N}\left[\frac{1}{N-1} \sum _{j\ne i}^{N-1}\theta (r-r_{ij}) \right] ^{q-1}
\noindent
\end{equation}
For $q=2$, the quantity contained inside the brackets in Eq. (\ref{EQ12}) is the probability of two randomly chosen points (here our passive tracers) being within a distance $r$ of one another.  For $q \ge 3$, the bracketed function is the number of q-tuplets of points (particles) whose pairwise distance is less than $r$.  Here $N$ are the total number of tracer particles (forced to be constant), $\theta$ is the heaviside step function, and $r_{ij}$ is the distance between particle $i$ and $j$. This algorithm for determining the spectrum of fractal dimensions was given by Hentschel and Proccacia \cite{Procaccia1983}.

To calculate $D_q$, the $log$ of the correlation sum  (Eq. (\ref{EQ12})) is plotted
versus the $log$ of $r$. The range of $r$ over
which the plot is a straight line is the scale-free (or scaling) region. The slope of the line dlog($C_q(r)$)/dlog(r)
is the value of $D_q$. The abscissa is $r$ in units of the dissipative scale $\eta$ (Table \ref{table}).

\begin{figure}[htbp]
\centering
\includegraphics[width=3.75 in]{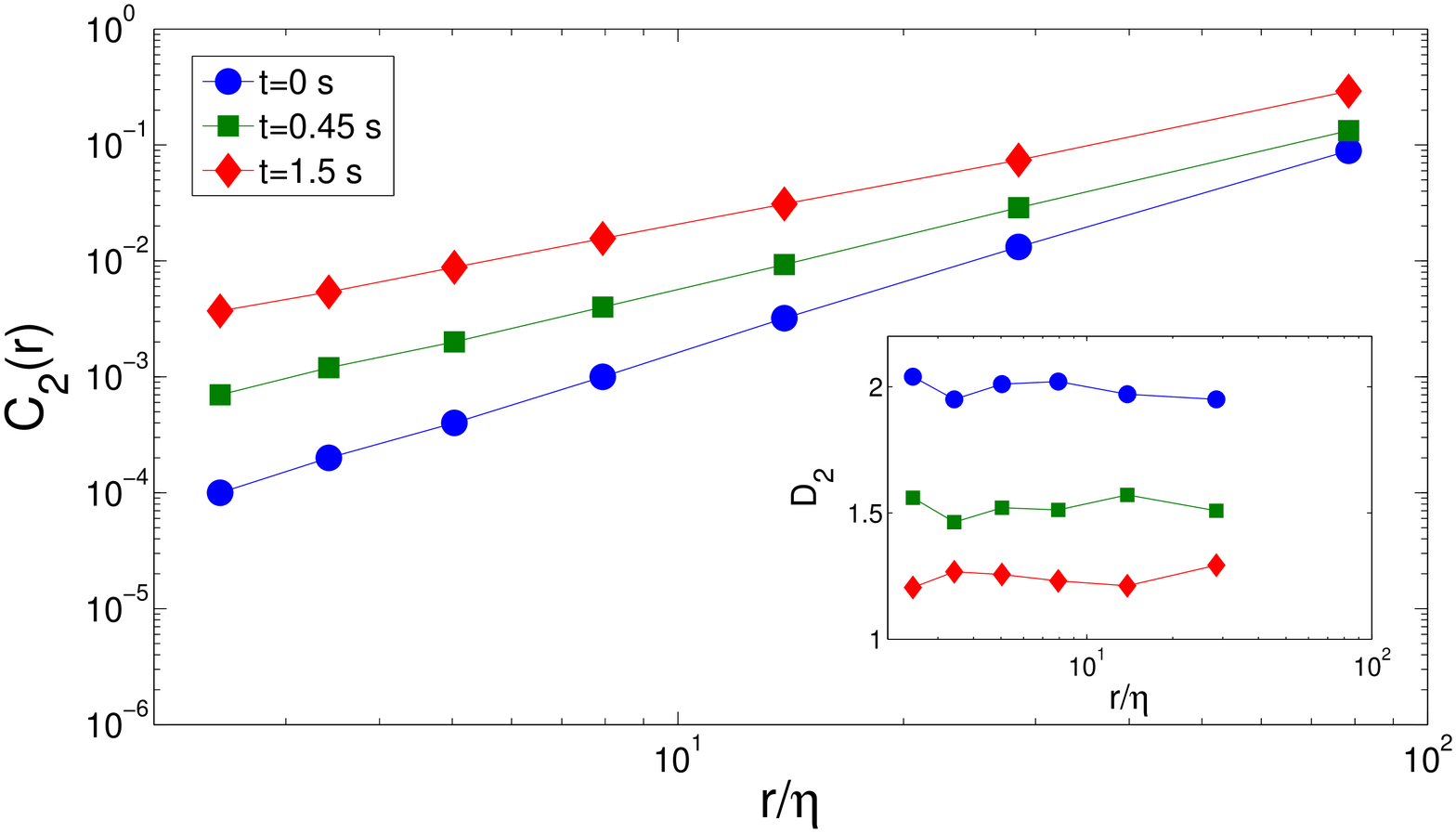}
\caption{\small{Correlation functions $C_2(r)$ averaged over several $Re_\lambda \simeq 160$ for various times in the experiment.  The interval of $r$ that exhibits scale-free behavior is $r/\eta \simeq 10^{2}$ to $10^{0.5}$. The inset shows the value $D_2$ calculated from the main figure.  Note that these scales are in the inertial range of the flow (Table I).}}
\label{D2_scaling}
\end{figure}

At $t=0$ the particles are uniformly distributed, $D_q(0)=2$. Figure \ref{D2_scaling} is a plot of $log(C_2(r))$ versus $log(r)$ for $t=0$ and subsequent times.  The scaling range is seen to span the interval $10^{0.5} <  r/\eta <10^{2}.$ The ratio of the integral scale $l_0$ to the dissipative scale $\eta$ is $l_0 / \eta \simeq 70 $. The initial homogeneous particle separation is roughly $\eta$. At $t=0$, $D_2(t=0)\simeq 2$ for $3 < r/ \eta < 70 $.  For scales $r/ \eta < 2$, $D_2(t=0) \simeq 0$, indicating the point-like nature of the Lagrangian tracer particles at scales less than initial tracer spacings. Thus, to ensure that the $D_q$ are defined such that $D_q > 0$ for all times $t$, the only measurements presented here are for  $r/ \eta > 2$.

\begin{figure}[htbp]
\centering
\includegraphics[width=3.75 in]{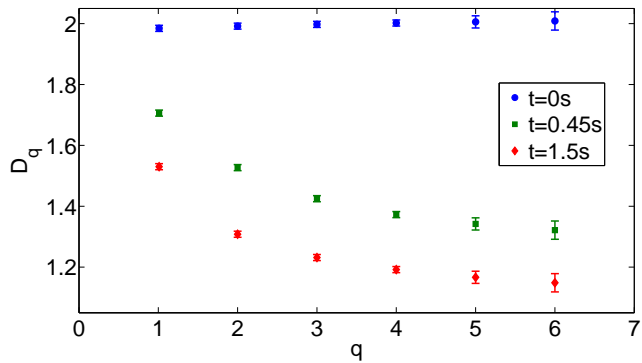}
\caption{\small{Range of values $D_q$ measured for $Re_\lambda = 169$ for various times in the experiment. These results indicate a multi-fractal distribution of the particles for the inertial range of the flow after the particle distribution evolves from a homogeneous distribution at $t=0$ s.}}
\label{Dq}
\end{figure}

Figure \ref{Dq} shows measurements of $D_q$ for $Re_\lambda \simeq 160$ for a range of $q=1$ to $6$ at various times.  The measurements at $t=1.5s$ are in the steady state (discussed below). The results indicate a multi-fractal distribution over what is usually considered to be the inertial range of the flow. However, other experiments \cite{Meneveau1996,Benzi1993} observe that the transition from the inertial to dissipative range occurs at spatial scales greater than $\eta$. Because of this, it is difficult to label the measurements here as being strictly inertial or dissipative. While the theory in \cite{Bec2004} predicts a multi-fractal particle distribution for compressible flows below the dissipative scale, no theory exists for the inertial range. The results in this work do agree qualitatively with the existence of inertial-range scaling seen in a numerical simulation \cite{Ducasse2008}. The following analysis of the time-evolution of the dimensions $D_q$ will focus on the so-called correlation dimension $D_2$.

The experiment was performed at several $Re_\lambda \simeq 150-170$, with an average $Re_\lambda \simeq 160$.  From each of these experiments, 17 time traces of the various $D_q(t)$ were measured starting with an initial homogeneous distribution.  These individual time traces were then used to produce an ensemble measurement of $D_q(t)$, which is analyzed subsequently. The total time spanned by the ensemble of experiments is approximately $25 s$, or roughly $60$ LETT's. For $t>1.5$, all of the $D_q$ measured fluctuate around a steady-state limiting value $D_q(\infty)$. For $D_2(t)$,  $D_2(\infty)=1.25$. Figure \ref{Dq} shows an example of $D_q(t)$ for $Re_\lambda = 169$ measured at various times $t<1.5s$, after which the dimensions saturate to values of $D_q(\infty)$. These values $D_q(\infty)$ are approximately those at $D_q(t=1.5s)$. The error bars reported in Figure \ref{Dq} are due to the error in the fits used to obtain the $D_q(t)$. Similar errors are found from the statistical ensemble averaging of the $D_q(t)$.

\begin{figure}[htbp]
\centering
\includegraphics[width=3.75 in]{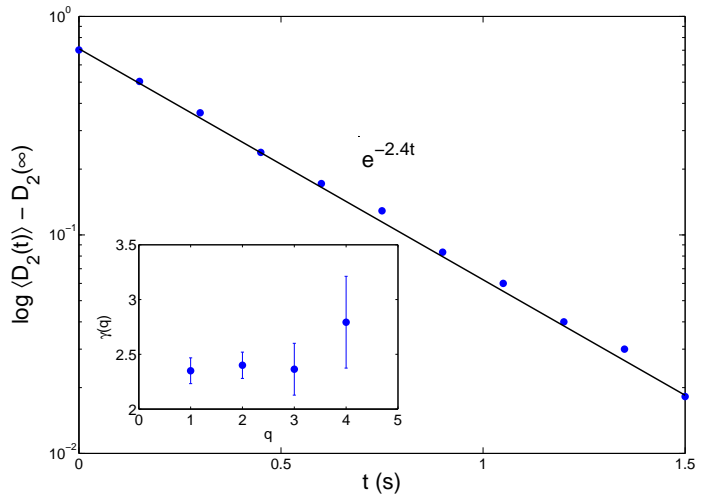}
\caption{\small{$D_2(t)$ averaged over several $Re_\lambda \simeq 160$. The result is an exponential decay of $D_2(t)$ from $D_2(0)=2$ to roughly 1.25 in approximately 1.5 s.  The solid line is a best fit to the data. The inset shows the decay exponent $\gamma(q)$ (discussed in text) as it varies with $q$.}}
\label{logD2}
\end{figure}

Our main result appears in Figure \ref{logD2}. This figure shows that $D_2(t)$ (ensemble averaged) decays at an exponential rate from $D_2(0)=2$ to $D_2(\infty)$: $D_2(t) - D_2(\infty) \propto exp(-\gamma t)$, where $\gamma=2.4 \pm 0.1 Hz$.
The decay time $\tau=1/ \gamma$ is $0.4s$, which is approximately one large-eddy turnover time $\tau _{0}$ (Table \ref{table}). This is the typical time for the largest "eddies" to significantly distort in a turbulent flow \cite{Frisch}. Several other $D_q(t)$ also experience an exponential decay from an initially homogeneous state $D_q(0)=2$ to their steady-state limiting value $D_q(\infty)$. The inset of Figure \ref{logD2} shows the exponent $\gamma(q)$ characterizing the exponential decay. One can see that $\gamma(q)$ is approximately independent of $q$ for $q=1..3$.  However, for $q>3$ the error in the measurement makes it difficult to predict that for large $q$, $D_q(t)$ will decay exponentially. It may also be worth noting that the value of $\gamma(q=1..3)$ is close to the value of the smallest Lyapunov exponent (but the largest in magnitude) measured in \cite{Boffetta2004} ($\lambda_2 \simeq -2 Hz$).  It remains to understand why this decay has exponential form.

\section{Summary}
A compressible system of particles in free-surface turbulence represents an instance of a chaotic attractor in real space. We study the time evolution of the low-order fractal dimensions $D_q$ for an ensemble of particles floating on a turbulent tank of water with the initial value of $D_q=2$ ($q$=1...6). The system evolves in a time of the order of the lifetime of the largest eddies, to a steady state where the measured $D_q(t)$ approach a value that is slightly greater than 1, implying the formation of string-like structures.  The correlation dimension $D_2(t)$ evolves exponentially as the steady-state is approached. It is not possible, so far, to deduce these observations from the Navier-Stokes equations in the inertial range of the compressible flow studied here.
\section{Acknowledgments}
{We thank B. Eckhardt and E. Ott for stimulating discussions.  This work was supported by the US National Science Foundation under grant No. DMR NSF 0604477. MMB carried out this work under the auspices of the National Nuclear Security Administration of the U. S. Department of Energy at Los Alamos National Laboratory under Contract No. DE-AC52-06NA25396.}
\appendix



\end{document}